\documentclass[
reprint,
superscriptaddress,
showkeys,
 amsmath,amssymb,
 aps,
]{revtex4-2}

\usepackage{graphicx}
\usepackage{dcolumn}
\usepackage{bm}
\usepackage{hyperref}
\usepackage{xcolor}
\usepackage{titlecaps}
\usepackage{tikz}
\usetikzlibrary{shapes}
\usepackage{cancel}
\usepackage{caption}
\usepackage[normalem]{ulem}
\usepackage{float}
\renewcommand{\figurename}{Fig.}

\begin{document}


\title{Differential Crosslinking and Contractile Motors Drive Nuclear Chromatin Compaction}

\author{Ligesh Theeyancheri}
\email{ligeshbhaskar@gmail.com}
\affiliation{Physics Department, Syracuse University, Syracuse, NY 13244 USA}
\author{Edward J. Banigan}
\email{ebanigan@mit.edu}
\affiliation{Institute of Medical Engineering \& Science and Department of
Physics, Massachusetts Institute of Technology, Cambridge, MA 02139, USA}
\author{J. M. Schwarz}
\email{jmschw02@syr.edu}
\affiliation{Physics Department, Syracuse University, Syracuse, NY 13244 USA}
\affiliation{Indian Creek Farm, Ithaca, NY 14850 USA}


\begin{abstract}
\noindent During interphase, a typical cell nucleus features spatial compartmentalization of transcriptionally active euchromatin and repressed heterochromatin domains. In conventional nuclear organization, euchromatin predominantly occupies the nuclear interior, while heterochromatin, which is approximately 50\% more dense than euchromatin, is positioned near the nuclear periphery.   Peripheral chromatin organization can be further modulated by the nuclear lamina, which is itself a deformable structure. While a number of biophysical mechanisms for compartmentalization within rigid nuclei have been explored, we study a chromatin model consisting of an active, crosslinked polymer tethered to a deformable, polymeric lamina shell. Contractile motors, the deformability of the shell, and the spatial distribution of crosslinks all play pivotal roles in this compartmentalization. We find that a radial crosslink density distribution, even with a small linear differential of higher crosslinking density at the edge of the nucleus, combined with contractile motor activity, drives genomic segregation, in agreement with experimental observations. This arises from contractile motors preferentially drawing crosslinks into their vicinity at the nuclear periphery, forming high-density domains that promote heterochromatin formation. We also find an increased stiffness of nuclear wrinkles given the preferential heterochromatin compaction below the lamina shell, which is consistent with instantaneous nuclear stiffening under applied nanoindentation. We conclude with the potential for experimental validation of our model predictions.  
\end{abstract}


\maketitle

\section{Introduction}

\noindent The cell nucleus plays a vital role in eukaryotic cells by enclosing and safeguarding genetic information encoded within chromatin polymer fibers composed of DNA and histone proteins~\cite{misteli2020self, kalukula2022mechanics}. Collectively, these polymer fibers constitute spatially partitioned chromatin states within the nucleus, broadly known as euchromatin and heterochromatin. Euchromatin is the gene-rich, transcriptionally active, and loosely packed form, while heterochromatin is gene-poor, transcriptionally inactive, and densely packed~\cite{solovei2016rule,dekker2024chromosome}. In conventional nuclear organization, euchromatin resides in the nuclear interior, while heterochromatin is localized at the nuclear periphery~\cite{solovei2009nuclear,solovei2016rule,dekker2024chromosome}, with a density difference of 50\% or more between the two~\cite{imai2017density,ou2017chromemt}.\\

\noindent Recent studies suggest several plausible explanations for spatial segregation of peripheral heterochromatin and interior euchromatin concentrated in the inner nuclear region. 
There is substantial evidence that compartmentalization of the genome can result from mesoscale phase separation~\cite{barbieri2012complexity,jost2014modeling,di2016transferable,larson2017liquid,strom2017phase,rao2017cohesin,nuebler2018chromatin,falk2019heterochromatin,gibson2019organization,wang2019histone,sanulli2019hp1,amiad2021live,adame2023regulation,mirny2019two}, with peripheral heterochromatin maintained by attractive interactions with the lamina~\cite{falk2019heterochromatin,amiad2021live,bajpai2021mesoscale,attar2024chromatin}. 
Besides affinity-mediated interactions, differences in the other physical properties of heterochromatin and euchromatin may also contribute to their partitioning. 
Nuclear elastography measurements indicate that heterochromatin may be almost four times stiffer than euchromatin~\cite{ghosh2021image}; 
simulations predict that the conventional nuclear chromatin organization can be obtained by increasing the bending stiffness of the heterochromatin relative to euchromatin~\cite{cook2009entropic,brunet2021physical,girard2024heterogeneous}. Entropic effects from nonspecific protein binding may further contribute to chromatin compaction, clustering, and compartmentalization~\cite{brackley2013nonspecific,brackley2016simulated}. In addition to equilibrium effects, a number of studies suggest that 
compartmentalization of the genome can result from \textit{nonequilibrium} mesoscale phase separation driven by heterogeneous  activity~\cite{ganai2014chromosome,grosberg2015nonequilibrium,smrek2017small,liu2018chain,das2022enzymatic, chaki2023polymer,mahajan2022euchromatin,hilbert2021transcription,brahmachari2024temporally,goychuk2023polymer}.\\

\noindent While a nonequilibrium phase separation approach may provide a basis for chromatin compartmentalization, as the cell nucleus is an out-of-equilibrium system,  chromatin and the nucleus also exhibit highly nontrivial dynamics. Specifically, various experiments observe micron-scale correlated movement of genomic regions over seconds~\cite{zidovska2013micron,shaban2018formation,khanna2019chromosome,barth2020coupling,shaban2020hi,eshghi2021interphase,locatelli2022dna}.  This correlated motion is reduced in the absence of ATP or by reduction of activity by motors, such as transcribing RNA polymerases~\cite{zidovska2013micron,bruinsma2014chromatin,shaban2018formation,shaban2020hi}. At the same time, the deformable lamina shell exhibits a broadly distributed power spectrum~\cite{chu2017origin}. It has been demonstrated that both the correlated chromatin motion and the power spectrum can be quantitatively understood with an active, crosslinked polymeric system confined within a deformable shell~\cite{liu2021dynamic}. 
These correlated motions are thought to drive dynamic nuclear shape deformations, such as wrinkles, bulges, and ``blebs'', chromatin-filled nuclear protrusions that are prone to rupture~\cite{berg2023transcription, prince2025transcriptional}. 
Given these findings, we ask to what extent the dynamic architecture of the cell nucleus may be explained by the physics of active polymers. Specifically, is motor activity combined with connectivity within chromatin sufficient to compartmentalize chromatin into different forms, particularly euchromatin and heterochromatin? \\

\noindent Here, we report progress towards answering this question to arrive at a simple, novel biophysical mechanism for chromatin compartmentalization based on motor activity and varying spatial profiles of chromatin crosslinking. 
Various molecular motors and enzymes operate throughout chromatin, including RNA polymerase II and cohesin \cite{zidovska2020self,bird2025cohesin}. Likewise, various DNA- and chromatin-binding proteins may act as crosslinkers by bridging chromatin sites. In particular, heterochromatin protein 1 (HP1) is a crucial mediator of chromatin crosslinking and can be found in both heterochromatin and euchromatin~\cite{canzio2011chromodomain,canzio2014mechanisms,machida2018structural,strom2021hp1alpha,sokolova2024structural,kumar2020heterochromatin}. HP1's diverse roles, including heterochromatin formation, gene silencing, telomere maintenance, DNA replication, and repair, suggest that the interactions of HP1 with neighboring protein complexes may play a significant role in regulating its ability to crosslink chromatin.  Therefore, the localization of other protein complexes, such as Cullin 4B (CUL4B), interacting with HP1 could lead to a spatial distribution of crosslinks within the nucleus~\cite{yang2015crl4b}. \\

\noindent With this biological context, our model incorporates simple extensile and contractile monopolar motor activities, representative of ATP-driven processes in the nucleus~\cite{bruinsma2014chromatin,saintillan2018extensile,liu2021dynamic,berg2023transcription}. We also implement chromatin crosslinks that mimic chromatin-binding proteins like HP1~\cite{canzio2011chromodomain,machida2018structural,strom2021hp1alpha,redding2021dynamic}. To capture the effects of differential HP1 localization in the nucleus, we vary the spatial distribution of crosslinks using uniform, linear, and inverted radial crosslink profiles. Additionally, our model accounts for the deformable nature of the nuclear envelope, as the lamina is soft and undergoes shape fluctuations~\cite{chu2017origin, patteson2019vimentin, liu2021dynamic, tang2023indentation}. 
We hypothesize that patterning of physical constraints within chromatin that regulate chromatin spatial organization and facilitate mechanoresponses and dynamic adaptation of the nucleus and genome to internal and external perturbations~\cite{tajik2016transcription,stephens2019chromatin,nava2020heterochromatin,strom2021hp1alpha,hsia2022confined,tang2023indentation,liang2024microtopography,attar2025peripheral,kalukula2022mechanics,miroshnikova2022mechanical}. 

\section{Model}
\begin{figure*}
\centering
	\includegraphics[width=0.98\linewidth]{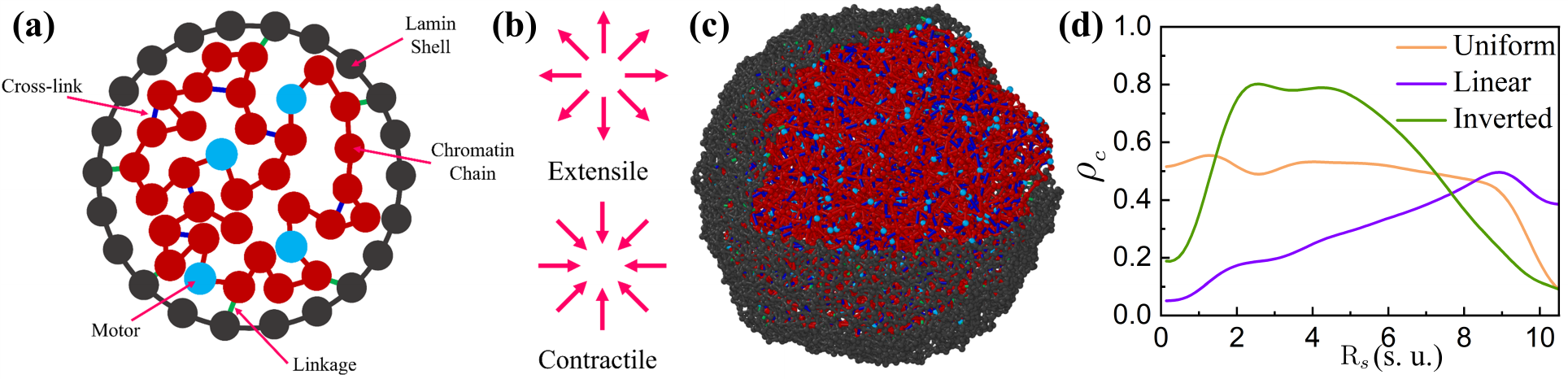}
\caption{\textit{3D computational model of chromatin in a deformable nucleus.} (a) A two-dimensional schematic of the model. (b) A schematic illustrating the two types of motors acting on the chromatin. (c) A Simulation snapshot. The chromatin polymer consists of linearly connected monomers, depicted in red, while the active chromatin subunits are shown in cyan. The lamina is made up of lamin subunits, represented in grey. (d) The radial distribution of crosslink density ($\rho_c$) as a function of nuclear radius $\text{R}_s$. All quantities are reported in simulation units (s. u.), where 1 s. u. of length = 1 $\mu m$.}\label{fig:model}
\end{figure*}
\noindent The chromatin-lamina system is modeled with the chromatin represented as a Rouse chain and the lamina as an elastic, polymeric shell, with linkages between the chain and the shell~\cite{stephens2017chromatin,banigan2017mechanics,liu2021dynamic}. More specifically, monomers in the lamina shell are connected by Hookean springs with spring constant $K$. In the initial preparation of the shell, a Fibonacci sphere with 5000 nodes is generated and 5000 identical monomers are placed at these nodes. The monomers connected by springs form a mesh with an average coordination number of 4.5. The monomers have soft-core repulsive interactions to model excluded volume. The chromatin fiber, on the other hand, is modeled by a Rouse chain with soft-core repulsion between each monomer, which has radius $r_p=0.43089$. A three-dimensional self-avoiding random walk in an FCC lattice is performed to generate the polymer with $N=5000$ monomers. Subsequently, the chromatin polymer is packed in a hard spherical shell before equilibration. The monomers of the shell have the same physical properties as the chain monomers in terms of size ($r_s$) and spring constant. \\ 

\noindent Initial configurations of the lamina and the chromatin confined within are generated using the following protocol~\cite{liu2021dynamic}. The spherical shell is gradually shrunk by moving its monomers inward, which in turn drives the chain monomers inward due to soft-core repulsions. Once the target radius is reached, the positions of the shell monomers are relaxed, and the spring lengths are updated to finalize the initial system configuration (Fig.~\ref{fig:sim_model}). With the initial shell radius, $\text{R}_s=10$, the resulting volume fraction is $\phi_{pack} \approx 0.4$  prior to further relaxation. We repeat this process and generate 50 initial configurations. 
We define the simulation units such that one unit of length corresponds to 1 $\mu$m, one unit of time is $0.5$ seconds, and the energy scale is $k_B T$ = $10^{-21}$~J, where $T = 300$~K (a complete list of simulation parameters is provided (Table~\ref{tab:parameters}) in the supporting information (SI)). \\ 
 
\noindent  We implement the excluded volume interactions using a repulsive, soft-core potential between any two monomers, $i$ and $j$, which are separated by a distance $|\vec{r}_{ij}|$, modeled using:
\begin{equation}
V_{\textrm{Ex}}(r_{ij}) = \begin{cases} \frac{1}{2} K_{\textrm{Ex}} \left(| r_{ij}| - \sigma_{ij}\right)^2 \hspace{5mm} \mbox{if } |r_{ij}|< \sigma_{ij}, \\
0 \hspace{35mm} \mbox{otherwise},
\end{cases}
\label{eq:excluded}
\end{equation}
where $\sigma_{ij} = r_{p_i} + r_{p_j}$. We also incorporate crosslinks in chromatin~\cite{strom2021hp1alpha}, tethering of chromatin to the lamina, and motor activity within the chromatin polymer (Fig.~\ref{fig:model}). We establish a total of $N_c$ crosslinks in the chromatin by adding springs between randomly selected pairs of polymer monomers, where the springs have the same stiffness, $K$, as those connecting adjacent monomers along the polymer chain. A crosslink may be established if the distance between the selected monomers is less than $r_{\text{link}} = 3r_p$. \\

\noindent To investigate how the distribution of crosslinks influences the spatial organization of chromatin within the nucleus, we consider three distinct (initial) crosslinking profiles: uniform, linear, and inverted (Fig.~\ref{fig:model}d). In the uniform profile, crosslinks are formed randomly between any $N_c$ pairs of monomers. The linear profile involves a linearly increasing crosslink density from the center of the nucleus toward the nuclear periphery. Conversely, in the inverted profile, the crosslink density increases linearly from the nuclear periphery toward the center of the nucleus. Crosslinks are assigned at the start of the simulation by dividing the nucleus into concentric spherical shells of equal volume and assigning a radial probability distribution for crosslink formation across these shells. For the $i^{th}$ shell, the probability of forming a crosslink is proportional to $\frac{r_i}{R_s}$ for the linear profile or $1 -\frac{r_i}{R_s}$ for the inverted profile, where $r_i$ is the shell center and $R_s$ is the nuclear center. This implies that the probability at the exact center is minimal due to the finite width of the innermost shell. Crosslinks are initialized once at the start of the simulation and remain topologically fixed throughout, although their spatial configuration evolves dynamically. The slope of the radial crosslinking profile remains fixed across simulations with different total numbers of crosslinks. This framework allows us to analyze how different spatial distributions and concentrations of crosslinks affect the three-dimensional arrangement of chromatin inside the nucleus. \\

\noindent A total of $N_L$ random polymer-shell linkages are introduced to model connections between chromatin and the nuclear lamina~\cite{kirby2018emerging,lionetti2020chromatin,attar2025peripheral}. In addition to thermal fluctuations, we incorporate motor activity by designating $N_m$ chromatin monomers as active~\cite{liu2021dynamic,berg2023transcription}. An active monomer exerts a force $\bm{F_a} = \pm f_{m} \hat{r}_{ij}$ on other monomers within a specified range. $f_{m}$ denotes the strength of the motor force. The motors can exert either extensile or contractile forces, pushing the monomers apart or drawing them together (Fig.~\ref{fig:model}b), analogous to other models that explicitly account for motor activity~\cite{bruinsma2014chromatin,saintillan2018extensile}. A motor stochastically unbinds from a monomer with characteristic turnover time, $\tau_m = 20$, after which the monomer becomes passive and the motor binds to another randomly selected chromatin monomer, which will then exert active forces. The timescale $\tau_m\approx 10$~s is comparable to the timescale of experimentally observed chromatin motions~\cite{zidovska2013micron,shaban2018formation}.\\ 

\noindent The system evolves through Brownian dynamics, governed by the overdamped equations of motion:
\begin{equation}
\xi  \bm{\dot{{r}}_{i}} = \bm{{F}_{Ex}} + \bm{{F}_{Sp}} + \bm{{F}_{Th}} + \bm{{F}_{{Ac}}} + \bm{f_n}(t),
\label{eq:langevineq}
\end{equation}
where $\bm{r_i}$ represents the position of the $i^{th}$ monomer at time $t$, and $\xi$ is the friction coefficient, which is related to the thermal Gaussian noise, $\bm{f_n}(t)$, via the fluctuation-dissipation theorem. $\bm{F_{Ex}}$ represents the force arising from excluded volume interactions, $\bm{F_{Sp}}$ accounts for the harmonic forces generated by polymer chain springs, chromatin crosslink springs, and chromatin-lamina linkage springs, and $\bm{F_{Th}}$ corresponds to the thermal forces, and $\bm{F_{Ac}}$ denotes the force due to the contractile or extensile motors. For the passive system, no motors are present, and the active force $\bm{F_{Ac}}$ is set to zero. We employ the Euler method for time integration, with a time step of $d\tau$ = $10^{-4}$, and a total simulation time of $\tau$ = $10^3$. 
To evaluate the structural properties in the steady state, we measure the radius of gyration, $R_g$, and average radius of the lamina shell, $\left < R_s \right >$. The system is considered to be in steady state when these quantities no longer exhibit significant changes over time (Fig.~\ref{fig:avg_rad_rg}).

\section{Results}

\noindent As we consider different types of motors and different crosslinking profiles, we also use a reference configuration, the chromatin distribution in the absence of both crosslinks and chromatin–lamina linkages, as well as in systems with lamina linkages but no crosslinks (Fig.~\ref{fig:den_ref}). Without chromatin-lamina linkages, chromatin accumulates at the center of the simulated nucleus, particularly in the scenario with contractile motors. This result suggests that chromatin-lamina linkages are essential to prevent excess accumulation of chromatin in the interior. In the following subsections, we systematically analyze how these molecular motors and crosslinks can establish conventional heterochromatin-euchromatin organization inside the nucleus.

\subsection{Radial chromatin distribution governed by motor activity and crosslink profile}
\begin{figure*}
\centering
	\includegraphics[width=0.9\linewidth]{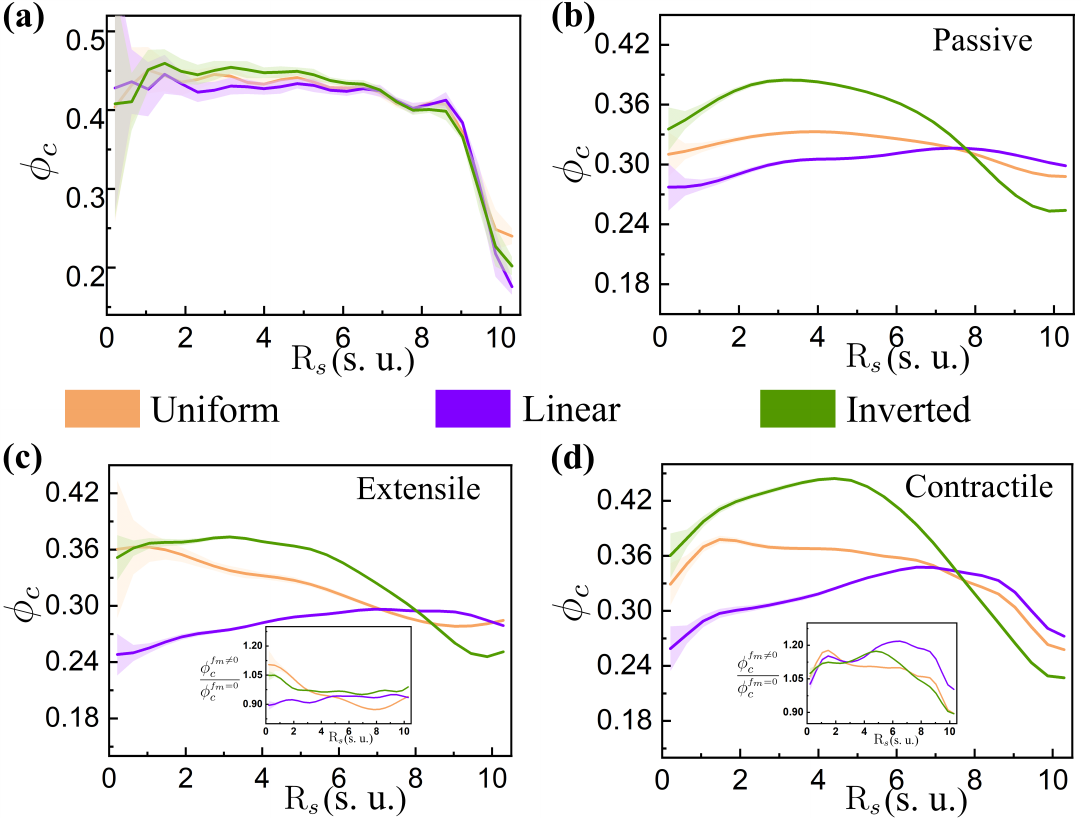}
\caption{\textit{Radial chromatin density shaped by motors and crosslink profiles.} Chromatin density, $\phi_c$, as a function of $\text{R}_s$. (a) Initial chromatin density at $t = 0$, and final chromatin density for (b) passive, (c) extensile, and (d) contractile motors, for different crosslink density profiles. The rescaled chromatin density, $\frac{\bm{\phi_c^{f_m \neq 0}}}{\bm{\phi_c^{f_m = 0}}}$, is shown in the insets of panels (c) and (d), highlighting motor-induced changes relative to the passive case. Each case corresponds to $N_c = 2000$ crosslinks and $N_L = 600$ chromatin-lamina linkages. All quantities are reported in simulation units (s. u.), where 1 s. u. of length = 1 $\mu m$ and 1 s. u. of time = 0.5 s.}\label{fig:Chden_profile} 
\end{figure*}
\noindent To quantify the effect of the crosslink distribution on the radial chromatin density, $\phi_c$, we compute $\phi_c$ by dividing the nuclear volume into concentric shells of equal width, counting the number of chromatin monomers in each shell, and normalizing by shell volume. Fig.~\ref{fig:Chden_profile} depicts $\phi_c$ $vs.$ radial position, $\text{R}_s$, for both initial configurations and steady states across passive, extensile, and contractile motor conditions with varying crosslink profiles. The initial chromatin distribution in the nucleus is uniform regardless of the crosslink density profile, as shown in Fig.~\ref{fig:Chden_profile}a. As the system evolves, chromatin spatial organization changes in response to the interplay of crosslinks and motor activity.\\

\noindent To distinguish the contribution of motor activity, we first examine the passive case without motors, where density profiles depend on the crosslink distribution. The uniform and linear crosslink profiles yield homogeneously distributed chromatin across $\text{R}_s$, while the inverted crosslink profile generates higher chromatin density in the nuclear interior (Fig.~\ref{fig:Chden_profile}b). \\

\noindent Motor activity alters the spatial pattern of chromatin density, depending on the crosslink distribution. Upon introducing extensile motors, chromatin localizes more toward the center in simulations with uniform and inverted crosslink density profiles. In contrast, with linear crosslink profiles, we observe a nearly uniform chromatin distribution as a function of $\text{R}_s$, with slightly elevated chromatin concentration at the periphery (Fig.~\ref{fig:Chden_profile}c). Notably, implementing contractile motors dramatically alters chromatin distribution across all three crosslink profiles. Uniform and inverted crosslink profiles lead to higher chromatin accumulation in the nuclear interior, with the inverted profile showing preferentially higher interior accumulation compared to the uniform profile. Strikingly, the linear crosslink profile results in significantly higher chromatin accumulation at the nuclear boundary (Fig.~\ref{fig:Chden_profile}d). 
This chromatin density profile resembles the conventional spatial distribution of chromatin observed in many differentiated cells~\cite{solovei2009nuclear,solovei2016rule} in that denser heterochromatin-like regions are near the nuclear periphery while less dense euchromatin-like chromatin is localized in the interior. \\

\noindent We also compare the spatial variation of chromatin density in the presence of motors to that in the corresponding passive system by examining the chromatin density scaled by the passive case, given as $\frac{\phi_c^{f_m \neq 0}}{\phi_c^{f_m = 0}}$ (insets in Fig.~\ref{fig:Chden_profile}c,d), where $\phi_c^{f_m \neq 0}$ and $\phi_c^{f_m = 0}$ denote the chromatin densities with and without (passive case) motor activity, respectively. This reveals that extensile motors generally lead to chromatin decompaction, whereas contractile motors result in significant chromatin compaction, particularly in the case of a linear crosslink profile. Our simulations demonstrate that motor activity, combined with nonuniformity in the spatial distribution of crosslinks, can alter the spatial distribution of chromatin. In particular, the synergistic action of contractile motors and a linearly increasing radial profile of crosslinks is sufficient to recapitulate the conventional nuclear architecture, suggesting a novel biophysical mechanism for the establishment of chromatin segregation patterns {\it in vivo}. \\

\begin{figure*}
\centering
	\includegraphics[width=0.98\linewidth]{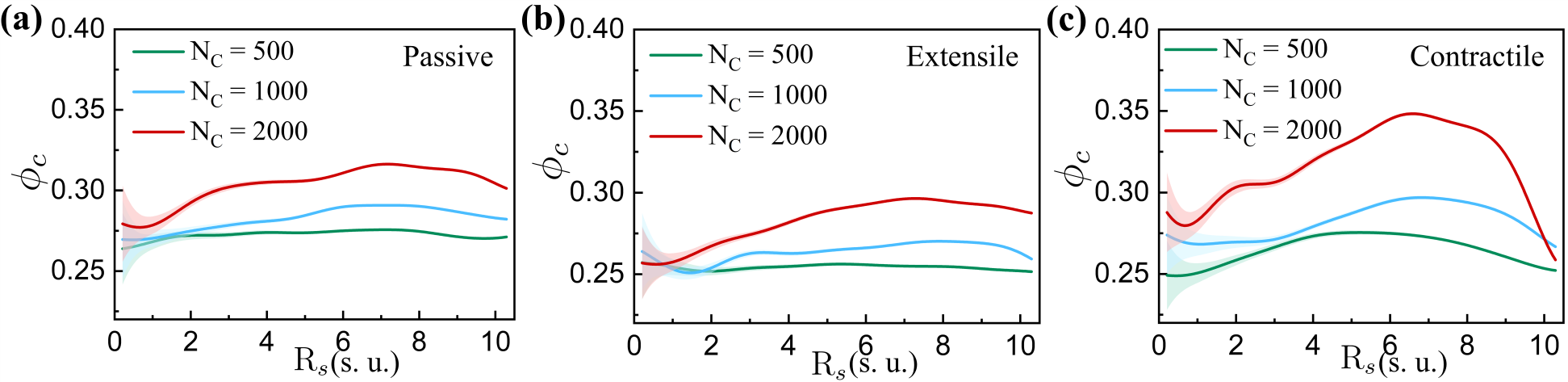}
\caption{\textit{Influence of motor activity and number of crosslinks on chromatin distribution.} Chromatin density, $\phi_c$, as a function of $\text{R}_s$ for (a) passive, (b) extensile, and (c) contractile motors, shown for different numbers of crosslinks ($N_c$) with a linear crosslink density profile. Results correspond to $N_L = 600$ chromatin crosslinks. All quantities are reported in simulation units (s. u.), where 1 s. u. of length = 1 $\mu m$ and 1 s. u. of time = 0.5 s.}\label{fig:Cden_xlinks} 
\end{figure*}

\noindent Subsequently, we examine how chromatin distribution changes for  different total number, $N_c$, of crosslinks in the system, while maintaining a  linear crosslink profile. Fig.~\ref{fig:Cden_xlinks} illustrates the effects across passive, extensile, and contractile motor conditions. Our findings reveal that chromatin compaction intensifies with increasing crosslink numbers, $N_c$, irrespective of motor activity. However, the magnitude of this effect varies markedly between different motor types. The passive and extensile cases (Fig.~\ref{fig:Cden_xlinks}a,b) demonstrate modest enhancement in chromatin density as $N_c$ increases, whereas the contractile case (Fig.~\ref{fig:Cden_xlinks}c) manifests substantially greater chromatin compaction. Among all three cases, extensile motors yield the least chromatin compaction because their activity locally expands chromatin.  The absence of motor activity enables intermediate levels of compaction with increasing $N_c$, as expected from entropic considerations. Contractile motors induce the highest degree of overall chromatin compaction.  Notably, the chromatin density is significantly enhanced at the nuclear periphery, reinforcing the conventional chromatin distribution pattern observed in experiments~\cite{solovei2009nuclear,solovei2016rule}. Additionally, increasing the number of chromatin-lamina linkages reinforces the spatial variation of chromatin density, leading to reduced density in the nuclear interior and accumulation near the periphery (Fig.~\ref{fig:Cden_LADs}). Our results establish that motor activity can focus spatial inhomogeneities in crosslinking into spatial patterning of chromatin polymer density. Contractile motor activity dramatically amplifies crosslink-dependent peripheral compaction, suggesting a general physical mechanism for regulating chromatin architecture through the interplay of molecular motors with the physical constraints on chromatin in the nucleus.

\subsection{Spatial segregation of euchromatin and heterochromatin contributes to overall chromatin density patterns}
\begin{figure*}
\centering
	\includegraphics[width=0.98\linewidth]{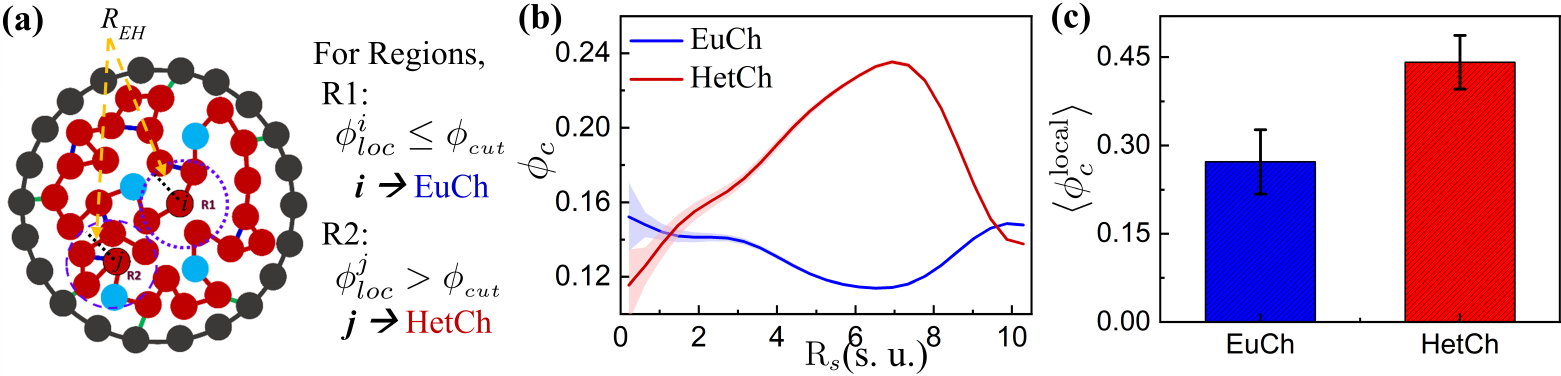}
\caption{\textit{Contractile motor-induced organization of euchromatin and heterochromatin.} (a) Schematic illustrating the categorization of chromatin monomers into euchromatin (EuCh) and heterochromatin (HetCh) based on a local density cutoff ($\phi^{i/j}_{loc}$). (b) Density of EuCh and HetCh as a function of $\text{R}_s$ and (c) average local density ($\langle \phi_c^{\text{local}} \rangle$) of EuCh and HetCh for a contractile motor with a linear crosslink density profile. The results shown correspond to $N_c = 2000$ crosslinks and $N_L = 600$ chromatin-lamina linkages. All quantities are reported in simulation units (s. u.), where 1 s. u. of length = 1 $\mu m$ and 1 s. u. of time = 0.5 s.}\label{fig:den_EH}
\end{figure*}

\noindent To quantify the observed patterns of chromatin architecture in our model, we sought to identify euchromatin-like and heterochromatin-like contributions to the spatial distribution of chromatin density. We focused on the simulations employing linear crosslink profiles with contractile motors, as this conditions generates chromatin distributions that most closely resemble conventional chromatin organization~\cite{solovei2009nuclear,solovei2016rule}. \\

\noindent To differentiate the radial chromatin density into euchromatin and heterochromatin components, we classified chromatin loci by local chromatin density. For each chromatin monomer $i$, we compute the local density $\phi_{loc}^i$ within a spherical volume of cutoff radius $R_{EH}$ centered on that monomer (Fig.~\ref{fig:den_EH}a). Monomers with $\phi_{loc}^i > \phi_{cut}$ (where $\phi_{cut}$ represents 95\% of the total chromatin density) are categorized as heterochromatin (HetCh), while the remaining monomers are classified as euchromatin (EuCh) (Fig.~\ref{fig:den_EH}a). The resulting density profiles of heterochromatin and euchromatin with a linear crosslink configuration are plotted as a function of the nuclear radius, $\text{R}_s$, in Fig.~\ref{fig:den_EH}b, revealing distinct, spatially segregated chromatin domains. We observe that heterochromatin is preferentially localized at the nuclear periphery, while euchromatin predominantly occupies the nuclear interior, a spatial arrangement that aligns with previously reported chromatin organization patterns~\cite{solovei2009nuclear,solovei2016rule}. 
In contrast to the contractile motor scenario, the passive and extensile cases exhibit a radial chromatin density profile that is predominantly shaped by euchromatin, with heterochromatin contributing less to the overall density throughout the nucleus (Fig.~\ref{fig:Cden_HetEuch}). This highlights that the interplay between crosslink distribution and motor activity could be a fundamental mechanism driving the characteristic spatial segregation and differential compaction of euchromatin and heterochromatin domains in the cell nucleus.\\

\noindent To further assess spatial chromatin compartmentalization, we compute the mean local densities of euchromatin and heterochromatin, $\langle \phi^c_{\text{local}} \rangle$. The average local densities are $\langle \phi^c_{\text{local}}(E) \rangle \approx 0.27$ for euchromatin and $\langle \phi^c_{\text{local}}(H) \rangle \approx 0.44$ for heterochromatin (Fig.~\ref{fig:den_EH}c). The resulting density ratio, $\frac{\langle \phi^c_{\text{local}}(H) \rangle}{\langle \phi^c_{\text{local}}(E) \rangle} \approx 1.6$, quantifies the degree of chromatin compaction, which is aligned with experimental reports of a ~50\% density increase in heterochromatin relative to euchromatin~\cite{imai2017density}. \\

\noindent We next characterize the dynamics of the euchromatin and heterochromatin using mean square displacement of monomers, defined as $\overline{\Delta r_i^{2}(\tau)} = \frac{1}{T^{\prime}-\tau} \int_{0}^{T^{\prime}-\tau} {\left[ \textbf{r}_i(t+\tau) - \textbf{r}_i(t)\right]}^2  dt$, where $\textbf{r}_i(t)$ denotes the position of the $i^{th}$ monomer at time $t$, $T^{\prime}$ is the total run time, and $\tau$ is the lag time (i.e., the time window over which displacements are averaged along a trajectory). To obtain the time-and-ensemble-averaged MSD, we compute the average, $\left\langle{\overline{\Delta r^{2}(\tau)}}\right\rangle  =  \frac{1}{N^{\prime}} \sum_{i=1}^{N^{\prime}}{\overline{\Delta r_i^{2}(\tau)}}$, where $N^{\prime}$ represents the number of independent trajectories. We observe that euchromatin displays slightly faster dynamics than heterochromatin (Fig.~\ref{fig:Cden_msd}), aligning with the trends noted in prior numerical~\cite{liu2018chain,shinkai2016dynamic} and experimental~\cite{minami2025replication,hathaway2012dynamics,nozaki2017dynamic} studies.

\subsection{Radial distribution of motors and the associated number of crosslinks per motor regulate chromatin compaction}
\begin{figure*}
\centering
	\includegraphics[width=0.98\linewidth]{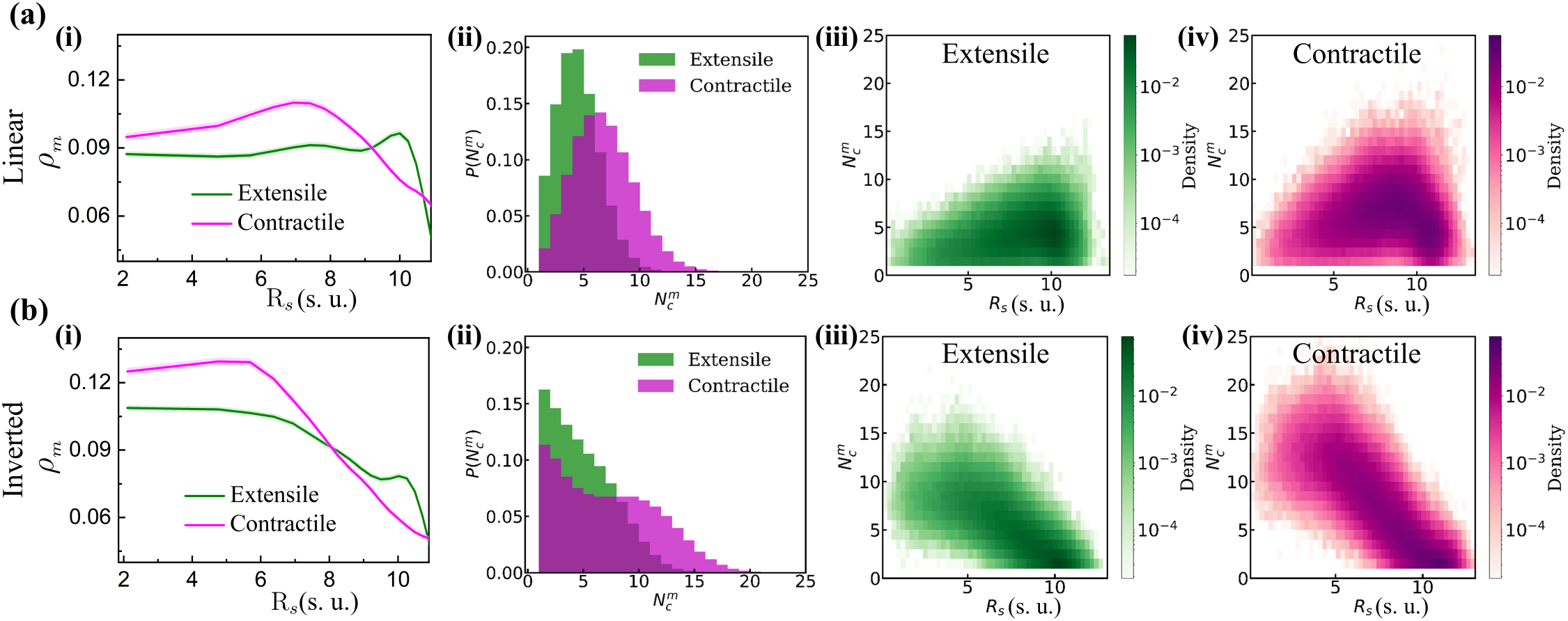}
\caption{\textit{Distribution of motor density and motor–crosslink associations for linear and inverted crosslink profiles.} Radial distribution of (i) motor density ($\rho_m$) and (ii) the average number of crosslinks per motor. Density color map of the number of crosslinks per motor ($N^m_c$) as a function of $\text{R}_s$ for  (iii) extensile and (iv) contractile motors. (a) Linear and (b) inverted crosslink profiles, each with subpanels (i–iv). The results shown correspond to $N_c = 2000$ crosslinks and $N_L = 600$ chromatin-lamina linkages. All quantities are reported in simulation units (s. u.), where 1 s. u. of length = 1 $\mu m$ and 1 s. u. of time = 0.5 s.}\label{fig:cmap_xlink_motor} 
\end{figure*}

\noindent The 3D structure of chromatin within the nucleus in our simulations arises from the interplay between mechanical constraints imposed by crosslinks and active, motor-driven processes that dynamically reorganize chromatin. To understand how motor forces and structural connectivity influence chromatin compaction and spatial arrangement, we characterize the radial distribution of motors and the density distribution of crosslinks that they establish throughout the nuclear volume.\\ 

\noindent We calculate the radial distribution of motors, $\rho_m$, and number of crosslinks per motor, $N^m_c$, followed by the density mapping of $N^m_c$ as a function of $R_s$ for linear crosslink profile. We compare this with the inverted crosslink profile, as these represent two extreme cases of spatial crosslink distribution within the nucleus, to identify the critical role of crosslink-motor interplay, as displayed in Fig.~\ref{fig:cmap_xlink_motor}. To estimate $\rho_m$, the nuclear volume is divided into concentric spherical shells of equal volume, starting from the center and extending to the nuclear boundary. At each time frame, the center of the nucleus is determined by calculating the mean position of the nuclear lamina particles, and the distance of each motor from this center is measured. The number of motors in each shell is then normalized by the volume of the respective shell, providing a measure of motor density as a function of radial distance, visualized in Fig.~\ref{fig:cmap_xlink_motor}(a, b)(i). To quantify $N^m_c$, for each motor monomer, we examine its local environment within a cutoff spherical volume of radius  $3r_p$ ($r_p$ is the radius of chromatin monomers) and count the number of crosslinked chromatin monomers within this sphere. This approach provides a direct measure of the average local crosslinking density, $\text{P}(N^m_c)$,  associated with each motor (Fig.~\ref{fig:cmap_xlink_motor}(a, b)(ii)). Furthermore, to understand the spatial dependence of crosslinks per motor, we map $N^m_c$ as a function of $R_s$ by binning motors according to their distance from the nuclear center and calculating the number of crosslinks per motor within each radial bin. This analysis reveals how crosslinking density varies across the nuclear volume and correlates with motor distribution patterns (Fig.~\ref{fig:cmap_xlink_motor}a,b(iii-iv)). \\

\noindent The radial distribution of motor density, $\rho_m$, reveals distinct spatial organization patterns for extensile and contractile motors depending on the crosslink density profile. For a linear crosslink profile, extensile motors show a relatively uniform density with a slight peripheral increase, whereas contractile motors peak in the outer regions and drop near the boundary (Fig.~\ref{fig:cmap_xlink_motor}a(i)). On the other hand, for an inverted crosslink profile, both motor types are enriched in the nuclear interior and decrease toward the periphery, with contractile motors exhibiting higher densities than extensile motors across $R_s$ (Fig.~\ref{fig:cmap_xlink_motor}b(i)). These motor distribution patterns correlate with the overall chromatin density distribution (Fig.~\ref{fig:Chden_profile}c,d), suggesting a strong coupling between motor localization and chromatin organization. Analysis of $N^m_c$ distributions uncovers striking differences between the two crosslink cases. In the linear case, contractile motors show a marked increase in crosslinks per motor compared to extensile motors, evident from the rightward shift of the $\text{P}(N^m_c)$ peak (Fig.~\ref{fig:cmap_xlink_motor}a(ii)). Conversely, the inverted crosslink case maintains a consistent peak location for both extensile and contractile motors, with the disparity between them nearly diminished (Fig.~\ref{fig:cmap_xlink_motor}b(ii)).\\

\noindent The density map further substantiates these observations, demonstrating that motors with the highest number of crosslinks are predominantly localized to the nuclear periphery in the linear crosslink profile when contractile motors are present (Fig.~\ref{fig:cmap_xlink_motor}a(iii, iv)). This pattern is opposite in the inverted crosslink configuration, where motors with higher crosslink densities are not only fewer in number but also distributed towards the nuclear interior (Fig.~\ref{fig:cmap_xlink_motor}b(iii, iv)). These findings underscore the biophysical mechanism by which the linear crosslink profile, coupled with contractile motor activity, promotes chromatin compaction and heterochromatin formation at the nuclear periphery. Contractile motors locally reinforce crosslink density, restricting chromatin mobility and facilitating the formation of dense chromatin domains. This process drives the segregation of heterochromatin at the periphery, while euchromatin remains enriched at the nuclear interior. \\
\begin{figure*}
\centering
	\includegraphics[width=0.95\linewidth]{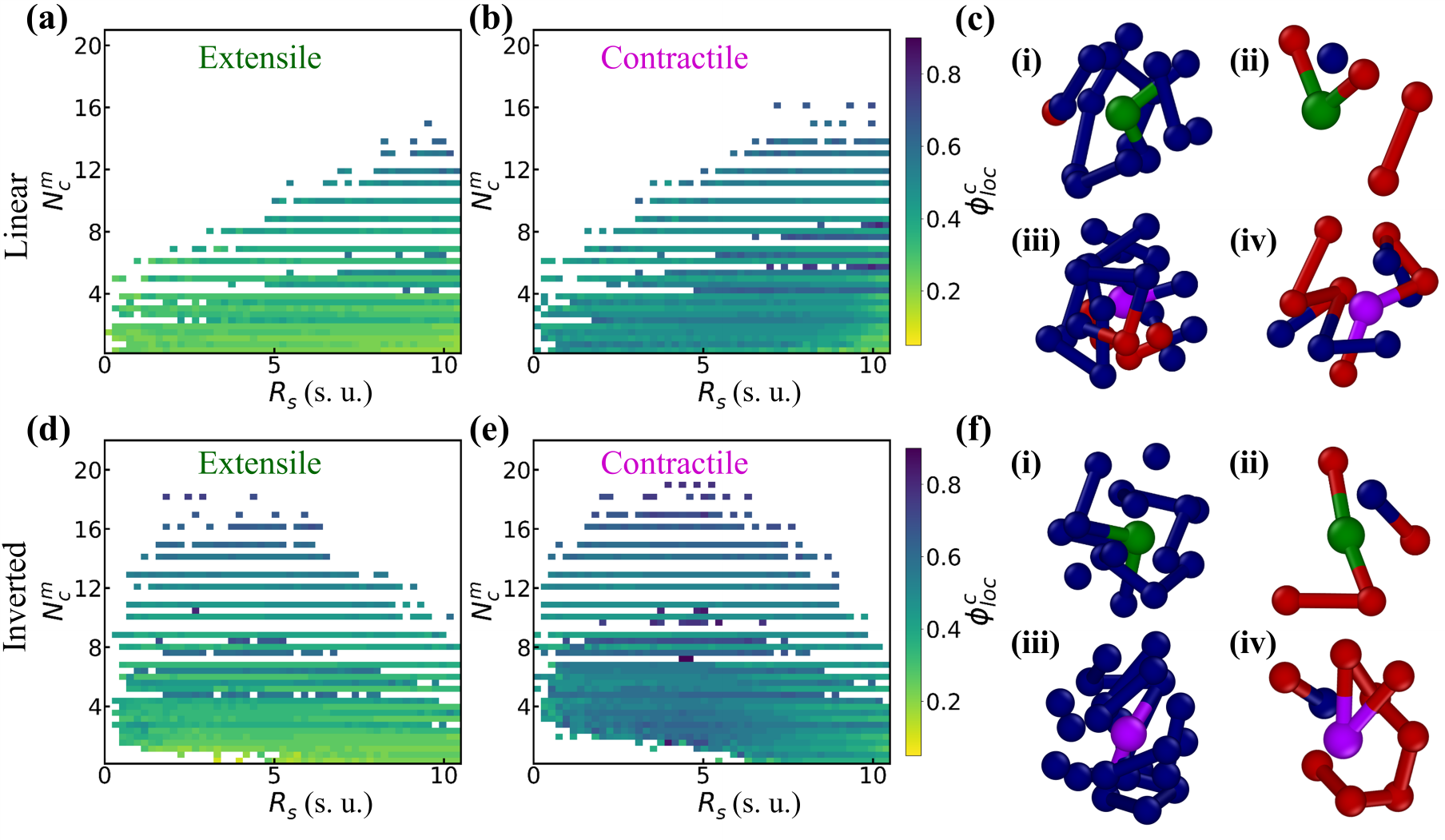}
\caption{\textit{Motor–crosslink coupling modulates local chromatin density for linear and inverted crosslink profiles.} Colormap of local chromatin density ($\phi^c_{loc}$) as a function of nuclear radius ($R_s$) and number of crosslinks per motor ($N^m_c$) for extensile and contractile motors for linear (top row: a, b) and inverted (bottom row: d, e) crosslink profiles. The color bar represents the normalized local chromatin density. Panels (c) and (f) show the local neighborhoods of an extensile motor (i, ii) and a contractile motor (iii, iv), with maximum and minimum crosslinks, respectively, for linear (top row) and reverse (bottom row) profiles. Magenta/green: contractile/extensile motors; blue: crosslinked chromatin; red: chromatin monomers. All quantities are reported in simulation units (s. u.), where 1 s. u. of length = 1 $\mu m$ and 1 s. u. of time = 0.5 s.}\label{fig:snapshot_density}
\end{figure*}

\noindent To examine the coupling between motor localization and chromatin density, we generated 2D colormaps of local chromatin density, $\phi^c_{\text{loc}}$, as a function of nuclear radius ($R_s$) and local crosslink number per motor ($N_c^m$). For each motor, $R_s$, $N_c^m$, and $\phi^c_{\text{loc}}$ were computed within a local neighborhood, and the resulting data were used to construct colormaps for both extensile and contractile motors for linear and inverted crosslink profiles. In the linear crosslink profile, chromatin density increases with $N_c^m$, particularly in the outer nuclear regions. Contractile motors exhibit stronger compaction--reflected by higher $\phi^c_{\text{loc}}$--towards the peripheral nuclear regions with increased $N_c^m$, consistent with their role in promoting local chromatin compaction (Fig.~\ref{fig:snapshot_density}(a, b)). For the inverted crosslink profile (bottom row), the density distribution shifts toward the nuclear interior, consistent with higher crosslink concentration near the center. Motors with higher $N_c^m$ again associate with locally compact chromatin, but this effect is more localized to the central region. This spatial dependence highlights the interplay between radial crosslink patterning, motor localization, and local chromatin organization (Fig.~\ref{fig:snapshot_density}(d, e)). \\

\noindent Consistent with the above findings, snapshots of the local neighborhoods surrounding contractile and extensile motors with maximum and minimum crosslinks for both linear (Fig.~\ref{fig:snapshot_density}c) and inverted crosslink profile (Fig.~\ref{fig:snapshot_density}f) strongly support our observations. Contractile motors with the maximum number of crosslinks exhibit the highest local crowding of chromatin (Fig.~\ref{fig:snapshot_density}c(iii), while extensile motors with minimal crosslinks display a significantly sparse local environment (Fig.~\ref{fig:snapshot_density}c(ii). Interestingly, contractile motors with the fewest crosslinks can form densely packed regions (Fig.~\ref{fig:snapshot_density}c(iv)), comparable to the local chromatin density observed around extensile motors with the highest number of crosslinks (Fig.~\ref{fig:snapshot_density}c(i)). Notably, this densification is more pronounced in the linear crosslink profile, highlighting the interplay of crosslink architecture and motor activity.

\subsection{Nuclear deformation enhances peripheral chromatin compaction}
\begin{figure*}
\centering
	\includegraphics[width=0.98\linewidth]{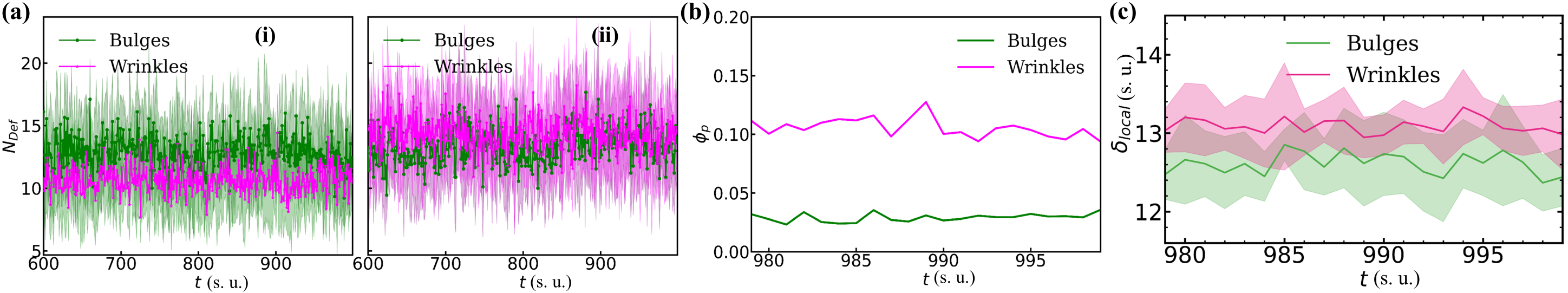}
\caption{\textit{Temporal profile of nuclear deformations and local properties of bulges and wrinkles with motor activity for a linear crosslink profile.} Time evolution of (a) the number of bulges/wrinkles ($N_{def}$) for (i) extensile and (ii) contractile motors, (b) chromatin density ($\phi_p$) at bulges/wrinkles, and (c) local fluctuations ($\delta_{local}$) characterizing bulges and wrinkle regions, all for a linear crosslink density profile. The results shown correspond to $N_c = 2000$ crosslinks and $N_L = 600$ chromatin-lamina linkages. All quantities are reported in simulation units (s. u.), where 1 s. u. of length = 1 $\mu m$ and 1 s. u. of time = 0.5 s.}\label{fig:BW_poly_stiffness}
\end{figure*}

\noindent Chromatin localization at the nuclear periphery is closely linked to changes in nuclear shape, as reported in earlier studies~\cite{poleshko2013human,    schreiner2015tethering,stephens2018chromatin,stephens2019chromatin,liang2024microtopography,tang2023indentation,berg2023transcription}. Dynamic deformations of the nuclear envelope, together with interactions between chromatin and the nuclear lamina, play a key role in regulating peripheral chromatin compaction. Hence, we quantitatively analyze the shape fluctuations of the soft nuclear lamina in our model. We characterize these fluctuations by measuring local geometric deviations from sphericity and computing the power spectrum of surface undulations (Fig.~\ref{fig:shape_fluct}). For each time frame, the center of mass of the nuclear lamina is first determined, along with the average radius ($\langle R_s \rangle$) by measuring the mean distance of all lamina monomers from this center. Local deformations are then classified into two distinct types: outward protrusions, referred to as bulges, and inward indentations, termed as wrinkles. To distinguish these deformation types on the three-dimensional lamina surface, regions containing groups of three or more lamina monomers are traced by calculating their radial deviations from $\langle R_s \rangle$. Deformations are classified based on a threshold of 10\% deviation from the average radius. Contiguous regions exceeding this threshold are identified as bulges if the deviation is positive, and as wrinkles if it is negative. To quantify the mechanical response of the deformed regions, we evaluate the local fluctuations for both bulges and wrinkles. Local positional fluctuations are quantified as the standard deviation of distances from each monomer in the bulges or wrinkles to their respective local center of mass.
The parameter $\delta_{\text{local}}$ is defined as the inverse of the squared fluctuations. Higher values of $\delta_{\text{local}}$ indicate regions with smaller deformations. Over long times, larger $\delta_{\text{local}}$ suggests greater local resistance to deformation, and therefore stiffer areas of the nuclear periphery, while lower values may correspond to more flexible regions.\\
    
\noindent In Fig.~\ref{fig:BW_poly_stiffness}, we plot the time evolution profile for the number of deformations, the chromatin density at these deformation sites, and the local fluctuations, $\delta_{\text{local}}$, associated with the bulges/wrinkles for the linear crosslink case. The temporal analysis of $N_{Def}$ reveals that contractile motors induce higher fluctuations and a greater number of wrinkles compared to extensile motors, which exhibit fewer deformations (Fig.~\ref{fig:BW_poly_stiffness}a). Furthermore, the polymer concentration at these deformation sites indicate a preferential localization of chromatin within the wrinkles rather than the bulges (Fig.\ref{fig:BW_poly_stiffness}b). This observation is supported by the inverse squared fluctuations over time, where wrinkles consistently exhibit larger $\delta_{\text{local}}$, suggesting higher stiffness compared to bulges (Fig.~\ref{fig:BW_poly_stiffness}c). The increased stiffness of the wrinkles is consistent with prior experimental investigations of the mechanical response of the cell nucleus to external forces, which found that nuclear envelope restructuring characterized by wrinkle unfolding is associated with instantaneous stiffening of the nucleus under applied indentation~\cite{tang2023indentation}. 

\section{Discussion}

\noindent We have analyzed how crosslink distribution and motor activity collectively influence chromatin spatial organization, promoting preferential peripheral heterochromatin formation and resulting in a spatially segregated architecture resembling conventional nuclear organization. This behavior is captured by our computational model, in which the lamina is represented as a soft deformable sphere and chromatin as a crosslinked polymer chain tethered randomly to the lamina, with dynamic motor proteins driving emergent chromatin compartmentalization. Each motor operates on a characteristic timescale, analogous to molecular motors observed \textit{in vivo}, after which it relocates to a different site along the chromatin chain. The interplay between spatial confinement imposed by crosslinks and the coherent, highly correlated local dynamics driven by motor proteins can either facilitate or hinder genomic interactions. The spatial distribution of crosslinks within the nuclear volume plays a critical role in regulating both the location and degree of chromatin compaction. Contractile motors further amplify this effect by promoting the formation of dense chromatin domains, particularly near the nuclear periphery. Additionally, wrinkling deformations of the nuclear lamina are correlated with accumulating chromatin at the boundary, which could self-reinforce peripheral organization.\\

\noindent Our study of chromatin organization using uniform, linear, and inverted spatial crosslink profiles reveal that a linear crosslink pattern, combined with contractile motor activity, leads to differential chromatin density--euchromatin predominantly localizes to the interior, while heterochromatin accumulates at the periphery. A local density-based classification approach, which maps individual densities of heterochromatin and euchromatin within the nucleus, further verifies this spatial compartmentalization, indicating an approximate 50\% difference in density between these two, consistent with experimentally reported values~\cite{imai2017density,ou2017chromemt}. The dynamic motors employed in our \textit{in silico} approach reveal that contractile motors markedly elevate the number of crosslinks per motor in their vicinity, resulting in strongly constrained local chromatin motion relative to extensile motors. This leads to the formation of high-density chromatin domains and mediating heterochromatin accumulation. Notably, this effect is most pronounced with the linear crosslink profile, where contractile motors with an increased number of crosslinks nearby are predominantly concentrated at the nuclear periphery, facilitating peripheral heterochromatin formation. In contrast, uniform and inverted crosslink profiles exhibit fewer motors with a higher number of crosslinks in their respective local environments, and when such motors are present, they are only marginally localized near the periphery. \\

\noindent The soft nuclear lamina exhibits shape fluctuations, as reported in previous studies~\cite{chu2017origin,patteson2019vimentin,liu2021dynamic,liang2024microtopography,tang2023indentation,talwar2013correlated}, and these deformations can give rise to the formation of bulges and wrinkles~\cite{tang2023indentation}. Wrinkles formed in the presence of contractile motors exhibit greater stiffness than bulges, which is directly correlated with a higher local concentration of chromatin near the wrinkles. 
Taken together, these findings suggest that a complete phase segregation into distinct low- and high-density chromatin regions is not strictly necessary to produce the conventional pattern of differential chromatin density observed in the nucleus. 
More importantly, our study indicates that local spatial effects driven by crosslink arrangement and motor activity are sufficient to establish the characteristic positioning of heterochromatin and euchromatin. \\

\noindent To experimentally validate our model’s predictions on the role of spatially patterned crosslinks and motor activity in shaping nuclear chromatin organization, we propose combining high-resolution chromatin imaging with genetic perturbations. Crosslink distribution can be visualized via immunofluorescence of endogenous proteins, such as HP1$\alpha$/$\gamma$, that crosslink chromatin and mark heterochromatin-associated domains~\cite{imai2017density,bintu2018super}. Crosslink density can be modulated by tuning HP1 expression levels: overexpression could promote chromatin crosslinking via HP1 oligomerization espeically in H3K9me3-enriched regions, while depletion or mutation would reduce crosslink formation~\cite{brasher2000structure, strom2017phase, larson2017liquid, strom2021hp1alpha}. Spatially patterned crosslinking could potentially be genetically engineered through synthetic recruitment of HP1 to defined genomic loci~\cite{ayyanathan2003regulated,verschure2005vivo} or optogenetically induced clustering~\cite{shin2017spatiotemporal,erdel2020mouse}, enabling spatial modulation of local crosslink density. Motor activity may be experimentally modulated by altering ATP availability or treating with inhibitors of transcription and other active processes~\cite{zidovska2013micron}. Locus tracking~\cite{khanna2019chromosome,gabriele2022dynamics} can assess the dynamic impact of crosslinks and motors. Together, these approaches would provide a versatile framework to test the mechanistic hypotheses emerging from our simulations.

\section{Acknowledgments}
\noindent This research was supported in part through computational resources provided by Syracuse University. EJB acknowledges support from the NIH Common Fund 4D Nucleome Program (UM1HG011536). JMS acknowledges funding support form from the National Science Foundation under grant DMR-2204312.

%
\clearpage
\onecolumngrid

\section*{Supporting Information} 

\renewcommand{\thefigure}{S\arabic{figure}}
\renewcommand{\figurename}{Fig.}
\setcounter{figure}{0}

\setcounter{section}{0}
\setcounter{subsection}{0}

\renewcommand{\thetable}{S\arabic{table}}
\setcounter{table}{0}
\captionsetup[table]{labelfont={bf},name={Table}}

\renewcommand\thesection{\arabic{section}}
\renewcommand\thesubsection{\thesection.\arabic{subsection}}

\section{C\lowercase{omputational} M\lowercase{odel}}
\begin{figure}[h!]
\centering
	\includegraphics[width=0.6\linewidth]{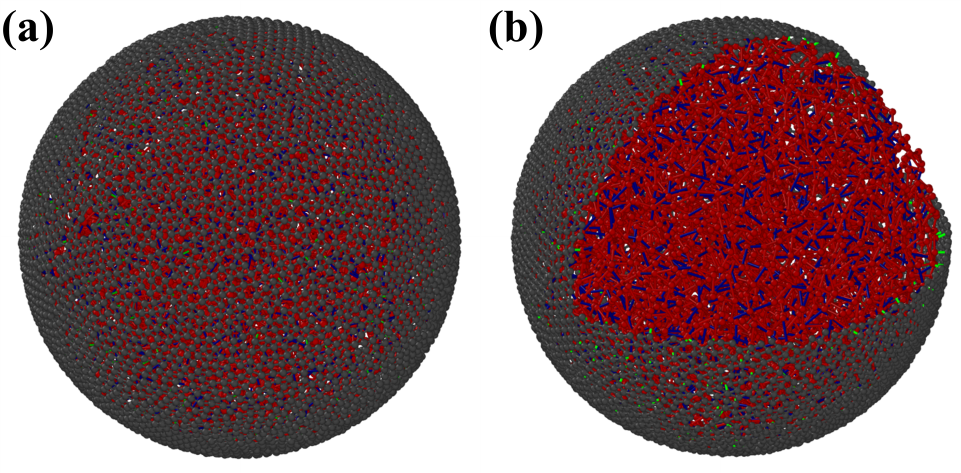} 
\caption{\textit{Snapshots of the initial configuration of the chromatin--lamina system.} (a) Initial configuration of chromatin confined within the nuclear lamina. (b) Crosslinked chromatin is shown with red beads; crosslinks are depicted in blue, and chromatin-lamina linkages are highlighted in green.}\label{fig:sim_model}
\end{figure}

\noindent The chromatin-lamina system is modeled with chromatin represented as a Rouse polymer chain and the nuclear lamina as an elastic, polymeric spherical shell. Linkages are formed between the chromatin and the lamina. The shell consists of 5000 monomers positioned using a Fibonacci sphere algorithm, with each pair of neighboring monomers connected by Hookean springs of spring constant $K$, forming a mesh with an average coordination number of 4.5. The chromatin is modeled as a Rouse chain with 5000 monomers ($N$), generated via a three-dimensional self-avoiding random walk on an FCC lattice. Each monomer has a radius $r_p = 0.43089$ and experiences soft-core repulsion to capture excluded volume effects. The lamina monomers share the same physical properties (size and spring strength) as the chromatin monomers. To initialize the system, the chromatin is confined within a spherical shell of radius $\text{R}s = 10$. The shell is gradually shrunk by moving lamina monomers inward, which compresses the chromatin via steric interactions. Once the desired radius is reached, spring lengths and monomer positions are adjusted to finalize the initial configuration (Fig.~\ref{fig:sim_model}). This process is repeated to generate 50 independent initial configurations. Given $r_p = 0.43089$, the packing fraction in the confined shell is approximately $\phi_{\text{pack}} \approx 0.4$ (details of simulation parameters are provided in Table~\ref{tab:parameters}).

\noindent Chromatin crosslinks are introduced by adding $N_c$ springs between randomly selected pairs of chromatin monomers, provided their mutual distance is less than $r_{\text{link}} = 3r_p$. The stiffness of the crosslink springs matches that of the polymer backbone.Motor activity is incorporated by designating $N_m$ chromatin monomers as active. Each active monomer exerts a force $\bm{F_a} = \pm f_m \hat{r}{ij}$ on nearby monomers, where $f_m$ denotes the motor force magnitude, and $\hat{r}{ij}$ is the unit vector between the interacting monomers. We consider two types of motors: extensile, which push monomers apart, and contractile, which draw them together. \\

\begin{table}[h]
\small
  \begin{tabular*}{0.6\textwidth}{@{\extracolsep{\fill}}lll}
    \hline
    Parameters & Numerical Value \\
    \hline
    Number of polymer monomers ($N$) & 5000  \\
    Number of shell particles ($N_s$) & 5000  \\
    Radius of a polymer monomer ($r_p$) & 0.43089  \\
    Radius of a shell particle ($r_s$) & 0.43089  \\
    Radius of the hard shell ($R_s$) & 10 \\
    Packing fraction ($\phi_{pack}$) & 0.4 \\
    Number of crosslinks ($N_c$) & 0, 500, 1000, 2000 \\
    Number of linkages ($N_L$) & 0, 50, 250, 600 \\
    Number of motors ($N_m$) & 500 \\
    Magnitude of motor force ($f_{m}$ ) & 10 \\
    Motor turnover time scale ($\tau_m$) & 20 \\
    Harmonic spring constant ($K$) & 140 \\
    Excluded volume strength ($K_{Ex}$) & 140 \\ 
    Damping ($\xi$) & 1 \\
    Simulation timestep ($d\tau$) & $10^{-4}$ \\
    Thermal Energy ($K_BT$) & 1 \\
    Diffusion constant ($D$) & 1 \\
    Cutoff used for crosslinks ($r_{\text{link}}$) & 3$\times$$r_p$ \\
    Motor radius ($r_{\text{motor}}$) & 1.5$\times$$r_p$ \\   
    \hline
  \end{tabular*}
    \caption{Parameters used in the simulations.}
  \label{tab:parameters}
\end{table}

\section{S\lowercase{imulation} R\lowercase{esults}}

\subsection{Radius of gyration of chromatin polymer and radial density distribution of chromatin monomers}

\begin{figure}[h!]
\centering
	\includegraphics[width=0.9\linewidth]{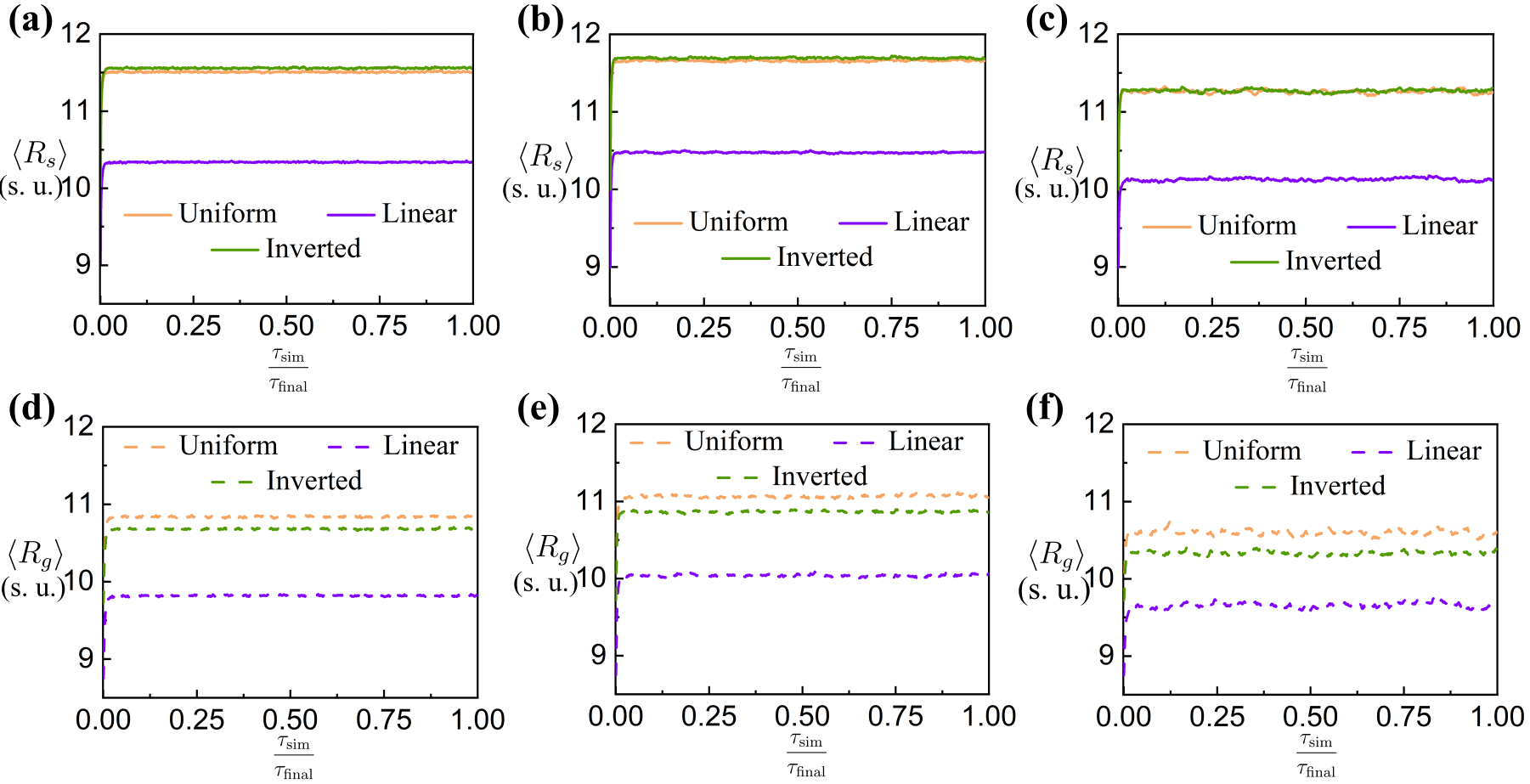} 
\caption{\textit{Temporal profile of nuclear radius and chromatin radius of gyration for passive, extensile, and contractile motors across different crosslink profiles.} Time evolution of the average radius of the nuclear lamina, $\langle R_s \rangle$ (solid lines, top row), and the chromatin radius of gyration, $\langle R_g \rangle$ (dashed lines, bottom row), for (a) passive, (b) extensile, and (c) contractile motor cases with different crosslink density profiles. The results shown correspond to $N_c = 2000$ crosslinks and $N_L = 600$ chromatin-lamina linkages. All quantities are reported in simulation units (s. u.), where 1 s. u. of length = 1 $\mu m$ and 1 s. u. of time = 0.5 s.}\label{fig:avg_rad_rg}
\end{figure} 

\noindent For a chromatin polymer, the radius of gyration is defined as $R_g = \sqrt{\frac{1}{N}\sum_{i=1}^N \left( \bm{r}_i - \bm{r}_{\text{com}} \right)^2 }$, where $N = 5000$ is the total number of monomers in the chain, $r_i$ is the position of $i^{th}$ monomer, and $r_\text{com}$ is the center of mass of the chromatin chain. The radius of the nuclear shell is fixed at $R_s = 10$ considering as an initial hard shell. The soft shell expands in response to thermal fluctuations and the active forces generated by the chromatin chain confined within the nucleus. Fig.~\ref{fig:avg_rad_rg} shows the average nuclear radius, $\langle R_s \rangle$ (top row, solid lines), and average radius of gyration, $\langle R_g \rangle$, of the chromatin chain (bottom row, dashed lines) as a function of simulation time. Following a short initial expansion, the radii of both the chromatin chain and the nuclear shell reach a plateau by 100$\tau$, indicating the onset of a steady-state configuration.

\subsection{Influence of chromatin-lamina linkages on chromatin spatial distribution}

\begin{figure}[h!]
\centering
	\includegraphics[width=0.65\linewidth]{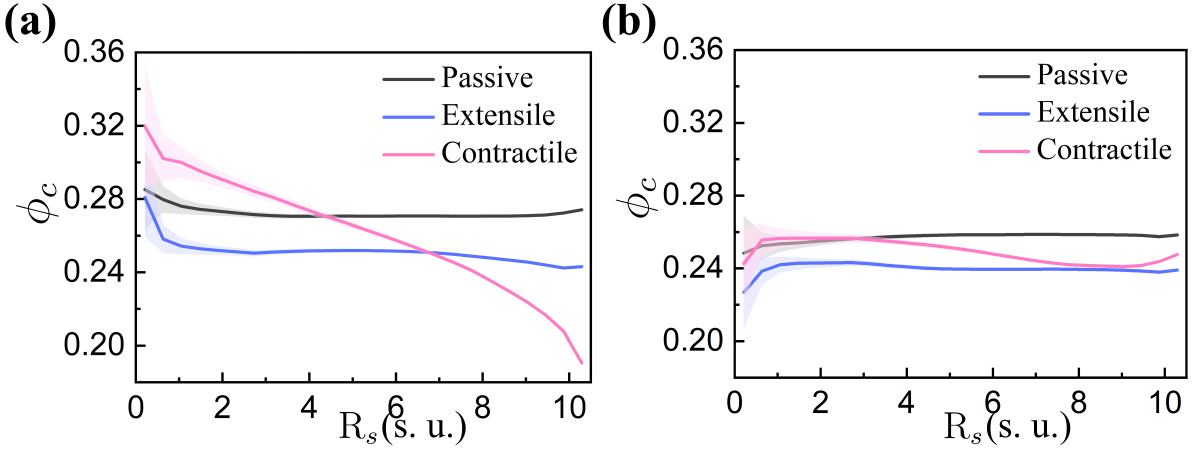} 
\caption{\textit{Chromatin density profiles for passive and motor cases without crosslinks and with chromatin--lamina linkage.} Chromatin density profiles, $\phi_c$, as a function of radial distance $\text{R}_s$ for passive, extensile, and contractile motor cases. (a) System without crosslinks or chromatin-lamina linkages ($N_c = 0$, $N_L = 0$). (b) System with chromatin-lamina linkages present but no crosslinks ($N_c = 0$, $N_L = 600$). All quantities are reported in simulation units (s. u.), where 1 s. u. of length = 1 $\mu m$ and 1 s. u. of time = 0.5 s.}\label{fig:den_ref}
\end{figure}

\noindent We consider a reference case in which both chromatin crosslinks ($N_c$) and chromatin–lamina linkages ($N_L$) are absent. The resulting radial chromatin distributions for passive, extensile, and contractile motor cases are shown in Fig.~\ref{fig:den_ref}a. In the absence of crosslinks and lamina linkages, the chromatin density remains largely uniform across the nuclear radius, $R_s$. A slight accumulation of chromatin is observed toward the nuclear interior, which becomes more pronounced in the presence of contractile motor activity (Fig.~\ref{fig:den_ref}a). \\

\begin{figure}[h!]
\centering
	\includegraphics[width=0.98\linewidth]{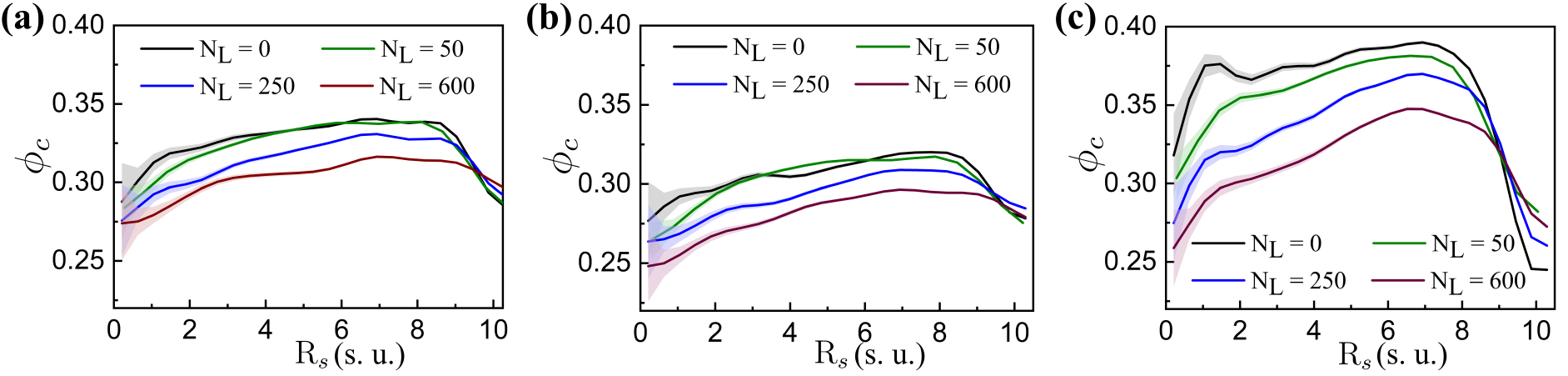} 
\caption{\textit{Chromatin distributions for passive, extensile, and contractile motors with varying chromatin--lamina linkages.} Chromatin density, $\phi_c$, as a function of $\text{R}_s$ for (a) passive, (b) extensile, and (c) contractile motors, shown for different chromatin-lamina linkage numbers ($N_L$) with a linear crosslink density profile. Results correspond to $N_c = 2000$ chromatin crosslinks. All quantities are reported in simulation units (s. u.), where 1 s. u. of length = 1 $\mu m$ and 1 s. u. of time = 0.5 s.}\label{fig:Cden_LADs}
\end{figure}
\noindent Next, we investigate how chromatin distribution is influenced by varying the number of chromatin–lamina linkages while keeping the crosslink profile linear and the total number of chromatin crosslinks fixed at $N_c = 2000$. Figure~\ref{fig:Cden_LADs} shows the resulting chromatin density profiles under passive, extensile, and contractile motor conditions. Both the passive and extensile cases yield nearly uniform radial density profiles, with extensile motor activity promoting chromatin decompaction relative to the passive and contractile cases (Fig.~\ref{fig:Cden_LADs}(a, b)). In contrast, contractile motors lead to significant chromatin compaction, resulting in a pronounced increase in density near the nuclear periphery. When no chromatin–lamina linkages are present, or when their number is low, the chromatin distribution remains relatively uniform, with slightly elevated density throughout the nucleus compared to cases with a higher number of linkages (Fig.~\ref{fig:Cden_LADs}c). However, increasing the number of linkages markedly enhances the spatial differentiation of chromatin density, with lower density in the nuclear interior and accumulation at the periphery (Fig.~\ref{fig:Cden_LADs}c).

\subsection{Spatial distribution and dynamics of euchromatin and heterochromatin}

\begin{figure}[h!]
\centering
	\includegraphics[width=0.98\linewidth]{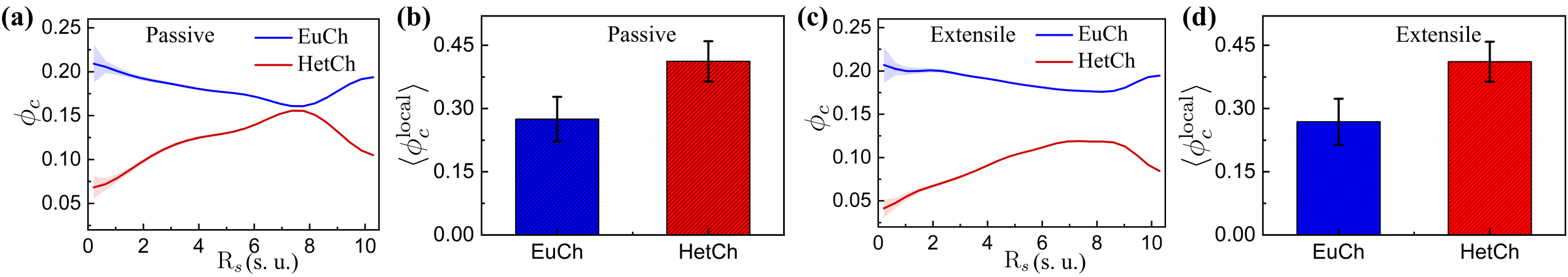} 
\caption{\textit{Euchromatin and heterochromatin density profiles for passive and extensile motor cases with a linear crosslink profile.} Density ($\phi_c$) of euchromatin (EuCh) and heterochromatin (HetCh) as a function of  $\text{R}_s$, along with the average chromatin density ($\langle \phi_c^{\text{local}} \rangle$), for passive (a, b) and extensile motor (c, d) cases, for a linear crosslink density profile. The results correspond to $N_c = 2000$ chromatin crosslinks and $N_L = 600$ chromatin-lamina linkages. All quantities are reported in simulation units (s. u.), where 1 s. u. of length = 1 $\mu m$ and 1 s. u. of time = 0.5 s.}\label{fig:Cden_HetEuch}
\end{figure}
\begin{figure}[h!]
\centering
	\includegraphics[width=0.45\linewidth]{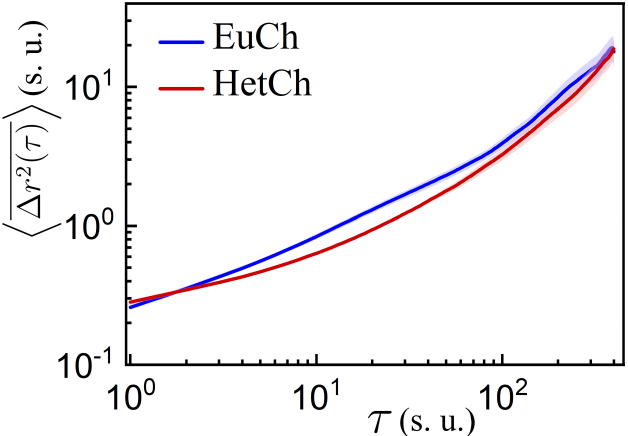} 
\caption{\textit{Dynamics of euchromatin and heterochromatin for contractile motors with a linear crosslink profile.} Mean square displacement $\left ( \langle{\overline{\Delta r^{2}(\tau)}}\rangle \right )$ of euchromatin (EuCh) and heterochromatin (HetCh) for contractile motors with a linear crosslink density profile. The results correspond to $N_c = 2000$ chromatin crosslinks and $N_L = 600$ chromatin-lamina linkages. All quantities are reported in simulation units (s. u.), where 1 s. u. of length = 1 $\mu m$ and 1 s. u. of time = 0.5 s.}\label{fig:Cden_msd}
\end{figure}
\clearpage
\subsection{Nuclear lamina shape fluctuations}

\begin{figure}[h!]
\centering
	\includegraphics[width=0.98\linewidth]{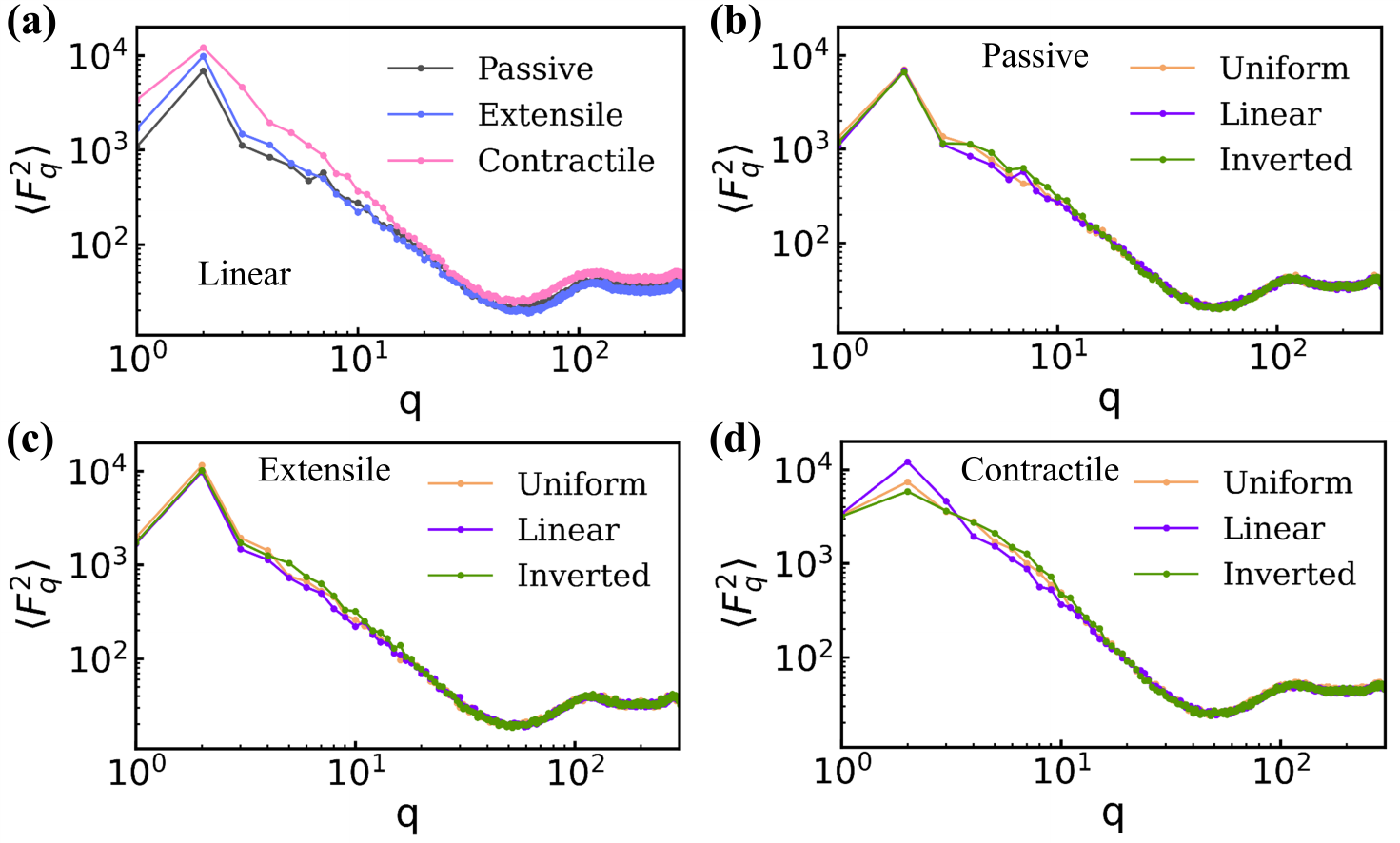} 
\caption{\textit{Nuclear shape fluctuations for different motor types and crosslink profiles.} Power spectrum of shell shape fluctuations for simulations with $N_c = 2000$ chromatin crosslinks and $N_L = 600$ chromatin-lamina linkages. (a) Different motor cases (passive, extensile, contractile) for a linear crosslink density profile. (b--d) Comparison of crosslink density profiles for (b) passive, (c) extensile, and (d) contractile motor cases.}\label{fig:shape_fluct}
\end{figure}

\noindent To characterize shape fluctuations of the shell, a random slab passing through the center is selected, and the coordinates of the shell monomers within the slab are projected onto its plane. The spatial deviations of these monomers from the average shell radius are then analyzed using a fast Fourier transform (FFT), with $F_q$ denoting the Fourier component of the deviation at wavenumber $q$. Fig.~\ref{fig:shape_fluct} presents the power spectrum of shape fluctuations for passive, extensile, and contractile motors with a linear crosslink profile, and for different crosslink density profiles corresponding to passive, extensile, and contractile motor cases. Contractile motors exhibit large-amplitude shape fluctuations, indicating more pronounced deformation of the nuclear lamina compared to extensile and passive cases (Fig.~\ref{fig:shape_fluct}a). However, for a given motor case, qualitatively similar trends are observed across different crosslink density profiles (Fig.~\ref{fig:shape_fluct}(b-d)).\\

\end{document}